\pdfoutput=1
\documentclass{article}
\usepackage{graphicx}
\PassOptionsToPackage{numbers, sort, compress}{natbib}
\usepackage[final]{neurips_2024}
\usepackage[utf8]{inputenc}
\usepackage[T1]{fontenc}
\usepackage{hyperref}
\usepackage{url}
\usepackage{booktabs}
\usepackage{amsfonts}
\usepackage{nicefrac}
\usepackage{microtype}
\usepackage[table]{xcolor}
\usepackage{array}

\title{Dimensions of Generative AI Evaluation Design}

\author{%
  P.~Alex Dow \quad Jennifer Wortman Vaughan \quad Solon Barocas \quad Chad Atalla \\\textbf{Alexandra Chouldechova \quad Hanna Wallach} \\
  Microsoft Research \\
  \texttt{\{alex.dow,\,jenn,\,solon,\,chad.atalla,\,alexandrac,\,wallach\}@microsoft.com} \\
}

\begin{document}

\maketitle

\begin{abstract}
There are few principles or guidelines to ensure evaluations of generative AI (GenAI) models and systems are effective. To help address this gap, we propose a set of general dimensions that capture critical choices involved in GenAI evaluation design. These dimensions include the evaluation setting, the task type, the input source, the interaction style, the duration, the metric type, and the scoring method. By situating GenAI evaluations within these dimensions, we aim to guide decision-making during GenAI evaluation design and provide a structure for comparing different evaluations. We illustrate the utility of the proposed set of general dimensions using two examples: a hypothetical evaluation of the fairness of a GenAI system and three real-world GenAI evaluations of biological threats.\looseness=-1
\end{abstract}

\section{Introduction}

Evaluating the capabilities and risks of generative AI (GenAI) models and systems is crucial for their successful development, deployment, and adoption.  Despite this, many would likely agree with \textit{New York Times} columnist Kevin Roose's recent characterization of the current state of GenAI evaluation as ``a~mess---a tangle of sloppy tests, apples-to-oranges comparisons and self-serving hype''~\citep{roose2024ai}.   
Even benchmarks, which have been viewed as the gold standard for measuring progress in GenAI model development, are coming under criticism.  Writing in the context of responsible AI, the 2024 AI Index Report identified ``a significant lack of standardization in responsible AI reporting'' wherein leading model developers test their models against different benchmarks, which the authors suggest ``may reflect existing benchmarks becoming quickly saturated, rendering them ineffective for comparison''~\citep{maslej2024ai}.  Furthermore, as NIST has observed, ``measuring risk at at an earlier stage in the AI
lifecycle may yield different results than measuring risk at a later stage,'' such as when a model becomes integrated into a system~\citep{nist2023artificial}.
 All in all, there are few principles or guidelines to ensure evaluations are effective, often making it unclear which approaches are suitable for which specific evaluation objectives. What GenAI evaluation lacks, in other words, is a systematic understanding of \textit{evaluation design}. To help address this gap, we propose a set of general dimensions that capture critical choices involved in GenAI evaluation design. Situating evaluations within these dimensions draws attention to important choices that might otherwise be overlooked---either when designing evaluations or evaluating them.\looseness=-1

Although 
several prior papers have provided useful decompositions for understanding GenAI evaluations, they are limited to specific types of evaluations, or evaluations of particular concepts. This includes work on disaggregated evaluations \citep{barocas2021designing}, sociotechnical safety evaluations \citep{weidinger2023sociotechnical}, evaluations with respect to the ``socio-technical gap'' \citep{liao2023rethinking}, human interaction evaluations \citep{ibrahim2024beyond}, and LLM benchmarks \citep{chang2024survey}.
We argue that key insights can be drawn by considering a set of general dimensions that are relevant to evaluations of \emph{any} GenAI model or system with respect to \emph{any} concept related to its capabilities or risks. The concept of interest, such as reasoning ability or stereotyping, is crucial, and numerous taxonomies of capabilities and risks provide different options to consider~\citep[e.g.,][]{blodgett2020languagetechnologypowercritical, blodgett2021sociolinguistically, liang2023holisticevaluationlanguagemodels, katzman2023taxonomizing}.
Just as important is the object of the evaluation---be it a model, a system, or a component thereof. However, once the concept and object have been specified, an evaluation designer is left with multiple critical choices about the most appropriate methodology for evaluating that object with respect to that concept.\looseness=-1

\section{Dimensions and Examples}

Building on \citeauthor{barocas2021designing}'s work on disaggregated evaluations~\citep{barocas2021designing},
we propose a set of general dimensions that capture critical choices involved in GenAI evaluation design and are 
relevant to evaluations of any GenAI model or system with respect to any concept. As shown in the left side of Table~\ref{tab:dims}, these dimensions include the evaluation setting, the task type, the input source, the interaction style, the duration, the metric type, and the scoring method.
We arrived at these dimensions by examining numerous evaluations, though they are not exhaustive and alternatives are likely possible.\looseness=-1

\begin{table}
    \centering
    \scriptsize
    \begin{tabular}{p{1cm} p{5.5cm} c >{\raggedright\arraybackslash}p{1.5cm} >{\raggedright\arraybackslash}p{1.5cm} >{\raggedright\arraybackslash}p{1.5cm}} 
        \toprule
        \textbf{Dimension} & \textbf{Description \& Example Values} & & \textbf{RAND} & \textbf{OpenAI} & \textbf{Google} \\ 
        \cmidrule{1-2} \cmidrule{4-6}
        \raggedright Evaluation setting & The setting in which the evaluation will take place, e.g., computer lab, wet lab, field test, production deployment & & \cellcolor{gray!50} Computer lab & \cellcolor{gray!50} Computer lab & \cellcolor{gray!50} Computer lab \\ 
        \cmidrule{1-2} \cmidrule{4-6}
        \raggedright Task type & The type of task involved in the evaluation, e.g., multiple choice questions, objective open-ended task, subjective open-ended task, real-world action facilitation & & \cellcolor{gray!50} Subjective open-ended & Objective open-ended & \cellcolor{gray!50} Subjective open-ended \\ 
        \cmidrule{1-2} \cmidrule{4-6}
        \raggedright Input source & The source of inputs to the model or system, e.g., human evaluator, human user/subject/expert, real world, AI & & \cellcolor{gray!50} Human subject & \cellcolor{gray!50} Human subject & Human evaluator \\ 
        \cmidrule{1-2} \cmidrule{4-6}
        \raggedright Interaction style & The style of interaction with the model or system, e.g., single turn, iterative & & \cellcolor{gray!50} Iterative & \cellcolor{gray!50} Iterative & Single turn \\ 
        \cmidrule{1-2} \cmidrule{4-6}
        \raggedright Duration & The duration of the evaluation, e.g., single session, longer duration, longitudinal & & Longer duration & \cellcolor{gray!50} Single sitting & \cellcolor{gray!50} Single sitting \\ 
        \cmidrule{1-2} \cmidrule{4-6}
        \raggedright Metric type & The type of metric, e.g., incidence, performance, feasibility, relative incidence/performance/etc. & & Relative feasibility & Relative performance & Incidence \\ 
        \cmidrule{1-2} \cmidrule{4-6}
        \raggedright Scoring method & The method for scoring outputs or behaviors, e.g., automated, human expert & & \cellcolor{gray!50} Human expert & \cellcolor{gray!50} Human expert & \cellcolor{gray!50} Human expert \\ 
        \bottomrule
    \end{tabular}
    \vspace{0.5cm}
    \caption{Left: Our proposed set of general dimensions that capture critical choices involved in GenAI evaluation design. Right: How three GenAI evaluations of biological threats, conducted by RAND~\citep{mouton2023operational}, OpenAI~\citep{patwardhan2024earlywarning}, and Google DeepMind~\citep{phuong2024evaluating}, might be situated within these dimensions. (Shaded cells for a dimension indicate evaluations that have the same value for that dimension.)}
    \label{tab:dims}
\end{table}

To illustrate the utility of this set of general dimensions, consider evaluating the fairness of a GenAI system. The concept of interest in such an evaluation might relate to events that compromise fairness, such as system outputs that stereotype, demean, or erase certain social groups, or 
to the negative fairness-related impacts of those events, such as reinforcing unjust social hierarchies. The particular concept chosen will affect the methodology that is most appropriate for evaluating the system.
The dimensions can help evaluation designers at this stage by drawing 
attention to various important choices that might otherwise be overlooked.  
If we are evaluating the system with respect to system outputs that stereotype, the evaluation might take place in a lab setting during a single session with inputs obtained from a variety of sources. However, if we are evaluating the system with respect to reinforcing unjust hierarchies, longitudinal field tests might be more appropriate with inputs provided by humans in iterative interactions with the system. In some cases, multiple complementary evaluations that represent different points on these dimensions may be needed to shed light on a particular concept.\looseness=-1

As a non-hypothetical example, consider three real-world GenAI evaluations of biological threats. The U.S. Department of Homeland Security (DHS)~\citeyearpar{dhs2024cbrn} noted that GenAI systems can lower barriers to entry for chemical, biological, radiological, and nuclear (CBRN) attacks and called for ``a standard framework... for pre-release evaluation and red teaming of AI models.'' Although no such standard framework currently exists, there is a growing body of work on evaluating GenAI models and systems with respect to CBRN threats. Table~\ref{tab:dims} shows our codings for how three recent GenAI evaluations of biological threats, conducted by RAND~\citep{mouton2023operational}, OpenAI~\citep{patwardhan2024earlywarning}, and Google DeepMind~\citep{phuong2024evaluating}, are situated within the dimensions. Several insights emerge from this:
(A) 
All three evaluations used human expert scoring, which may indicate that human judgment is essential for assessing complex biological threats, or it could suggest the need to explore alternative scoring methods.
(B) 
RAND and OpenAI's evaluations both involved human subjects generating inputs to a GenAI system with iterative interactions, while Google DeepMind's human evaluators worked with experts to create adversarial inputs for prompting a GenAI model with single-turn interactions. The former seems more consistent with the concerns \citet{dhs2024cbrn} voiced about lowering barriers to entry for human actors, while the latter might offer a more direct assessment of the model, avoiding the noise introduced by the variation in human subjects.
(C) While all three evaluations used different metric types, both RAND and OpenAI's evaluations used relative metrics, comparing outputs 
produced by multiple sets of human subjects, some with only access to the internet and others that also had access to a GenAI system.
This choice aligns with calls to assess the marginal risks of AI~\citep{kapoor2024societal}. In contrast, Google DeepMind's evaluation, which is still under development, focused on the incidence of ``problematic'' outputs~\citep{phuong2024evaluating}.\looseness=-1

\section{Discussion}

As the above examples illustrate, situating evaluations within our proposed set of general dimensions can draw attention to important choices that might otherwise be overlooked---either when designing evaluations or evaluating them. Specifically, the dimensions can guide decision-making during GenAI evaluation design, helping evaluation designers determine the most appropriate methodology for evaluating a particular GenAI model or system with respect to a particular concept. The dimensions can also provide a structure for comparing different evaluations. Given the critical role of evaluations in the successful development, deployment, and adoption of GenAI models and systems, we hope that our proposal encourages a more methodical and explicit approach to GenAI evaluation design.

\section{Broader Impacts}

By situating GenAI evaluations within a set of general dimensions, this paper seeks to address the current lack of principles or guidelines for GenAI evaluation design. We hope this will lead to a number of positive impacts, including improving evaluations, enhancing our collective understanding of the capabilities and risks of GenAI models and systems, and supporting their responsible development, deployment, and adoption. However, there are also potential negative impacts that must be considered and mitigated. First, there is a risk that the relatively small number of dimensions we proposed could give an illusion of simplicity, leading evaluation designers and other stakeholders to miss the necessary nuance involved in designing and conducting effective evaluations. Similarly, if these dimensions prove generally useful, there is a risk of overfocus, causing evaluation designers to overlook other important choices that are not captured by them. As we further expand on the work described in this paper, we can mitigate these potential negative impacts by articulating clear guidelines for their use and non-use, as well as encouraging the development of alternative structures for GenAI evaluation design.\looseness=-1

\section{Limitations}

It is possible that particular types of GenAI evaluations require other dimensions that we have not yet identified.  
In the future, we will seek 
to identify any gaps in our proposed set of general dimensions.\looseness=-1

To enhance the practicality of the dimensions, it would be beneficial to provide concrete examples of what can and cannot be learned from a GenAI evaluation given a particular choice. Such examples would help evaluation designers understand the implications of their choices and improve the overall utility of the dimensions. Additionally, to gain a deeper insight into the effectiveness of the dimensions, it is essential to apply them to a broader spectrum of evaluations across various concepts and objects. This expanded application would allow us to discern patterns and make further refinements. \looseness=-1

Finally, although we demonstrated how to examine existing GenAI evaluations, such as three real-world GenAI evaluations of biological threats, we touched only briefly (and hypothetically) on using these dimensions to design new evaluations. This gap highlights the need for further research.\looseness=-1

\bibliography{references}

\begin{thebibliography}{17}
\providecommand{\natexlab}[1]{#1}
\providecommand{\url}[1]{\texttt{#1}}
\expandafter\ifx\csname urlstyle\endcsname\relax
  \providecommand{\doi}[1]{doi: #1}\else
  \providecommand{\doi}{doi: \begingroup \urlstyle{rm}\Url}\fi

\bibitem[Barocas et~al.(2021)Barocas, Guo, Kamar, Krones, Morris, Vaughan,
  Wadsworth, and Wallach]{barocas2021designing}
Solon Barocas, Anhong Guo, Ece Kamar, Jacquelyn Krones, Meredith~Ringel Morris,
  Jennifer~Wortman Vaughan, W~Duncan Wadsworth, and Hanna Wallach.
\newblock Designing disaggregated evaluations of {AI} systems: Choices,
  considerations, and tradeoffs.
\newblock In \emph{Proceedings of the 2021 AAAI/ACM Conference on AI, Ethics,
  and Society}, pages 368--378, 2021.

\bibitem[Blodgett(2021)]{blodgett2021sociolinguistically}
Su~Lin Blodgett.
\newblock \emph{Sociolinguistically driven approaches for just natural language
  processing}.
\newblock PhD thesis, University of Massachusetts Amherst, February 2021.

\bibitem[Blodgett et~al.(2020)Blodgett, Barocas, III, and
  Wallach]{blodgett2020languagetechnologypowercritical}
Su~Lin Blodgett, Solon Barocas, Hal~Daumé III, and Hanna Wallach.
\newblock Language (technology) is power: A critical survey of "bias" in {NLP},
  2020.
\newblock URL \url{https://arxiv.org/abs/2005.14050}.

\bibitem[Chang et~al.(2024)Chang, Wang, Wang, Wu, Yang, Zhu, Chen, Yi, Wang,
  Wang, Ye, Zhang, Chang, Yu, Yang, and Xie]{chang2024survey}
Yupeng Chang, Xu~Wang, Jindong Wang, Yuan Wu, Linyi Yang, Kaijie Zhu, Hao Chen,
  Xiaoyuan Yi, Cunxiang Wang, Yidong Wang, Wei Ye, Yue Zhang, Yi~Chang,
  Philp~S. Yu, Qiang Yang, and Xing Xie.
\newblock A survey on evaluation of large language models.
\newblock \emph{ACM Transactions on Intelligent Systems and Technology},
  15\penalty0 (3):\penalty0 1--45, 2024.

\bibitem[{DHS}(2024)]{dhs2024cbrn}
{DHS}.
\newblock {D}epartment of {H}omeland {S}ecurity report on reducing the risks at
  the intersection of artificial intelligence and chemical, biological,
  radiological, and nuclear threats, April 2024.
\newblock URL
  \url{https://www.dhs.gov/sites/default/files/2024-06/24_0620_cwmd-dhs-cbrn-ai-eo-report-04262024-public-release.pdf}.
\newblock Accessed: 2024-09-16.

\bibitem[Ibrahim et~al.(2024)Ibrahim, Huang, Ahmad, and
  Anderljung]{ibrahim2024beyond}
Lujain Ibrahim, Saffron Huang, Lama Ahmad, and Markus Anderljung.
\newblock Beyond static {AI} evaluations: Advancing human interaction
  evaluations for {LLM} harms and risks.
\newblock \emph{arXiv preprint arXiv:2405.10632}, 2024.

\bibitem[Kapoor et~al.(2024)Kapoor, Bommasani, Klyman, Longpre, Ramaswami,
  Cihon, Hopkins, Bankston, Biderman, Bogen, Chowdhury, Engler, Henderson,
  Jernite, Lazar, Maffulli, Nelson, Pineau, Skowron, Song, Storchan, Zhang, Ho,
  Liang, and Narayanan]{kapoor2024societal}
Sayash Kapoor, Rishi Bommasani, Kevin Klyman, Shayne Longpre, Ashwin Ramaswami,
  Peter Cihon, Aspen Hopkins, Kevin Bankston, Stella Biderman, Miranda Bogen,
  Rumman Chowdhury, Alex Engler, Peter Henderson, Yacine Jernite, Seth Lazar,
  Stefano Maffulli, Alondra Nelson, Joelle Pineau, Aviya Skowron, Dawn Song,
  Victor Storchan, Daniel Zhang, Daniel~E. Ho, Percy Liang, and Arvind
  Narayanan.
\newblock On the societal impact of open foundation models.
\newblock \emph{arXiv preprint arXiv:2403.07918}, 2024.

\bibitem[Katzman et~al.(2023)Katzman, Wang, Scheuerman, Blodgett, Laird,
  Wallach, and Barocas]{katzman2023taxonomizing}
Jared Katzman, Angelina Wang, Morgan Scheuerman, Su~Lin Blodgett, Kristen
  Laird, Hanna Wallach, and Solon Barocas.
\newblock Taxonomizing and measuring representational harms: A look at image
  tagging.
\newblock In \emph{Proceedings of the AAAI Conference on Artificial
  Intelligence}, pages 14277--14285, 2023.

\bibitem[Liang et~al.(2023)Liang, Bommasani, Lee, Tsipras, Soylu, Yasunaga,
  Zhang, Narayanan, Wu, Kumar, Newman, Yuan, Yan, Zhang, Cosgrove, Manning,
  Ré, Acosta-Navas, Hudson, Zelikman, Durmus, Ladhak, Rong, Ren, Yao, Wang,
  Santhanam, Orr, Zheng, Yuksekgonul, Suzgun, Kim, Guha, Chatterji, Khattab,
  Henderson, Huang, Chi, Xie, Santurkar, Ganguli, Hashimoto, Icard, Zhang,
  Chaudhary, Wang, Li, Mai, Zhang, and
  Koreeda]{liang2023holisticevaluationlanguagemodels}
Percy Liang, Rishi Bommasani, Tony Lee, Dimitris Tsipras, Dilara Soylu,
  Michihiro Yasunaga, Yian Zhang, Deepak Narayanan, Yuhuai Wu, Ananya Kumar,
  Benjamin Newman, Binhang Yuan, Bobby Yan, Ce~Zhang, Christian Cosgrove,
  Christopher~D. Manning, Christopher Ré, Diana Acosta-Navas, Drew~A. Hudson,
  Eric Zelikman, Esin Durmus, Faisal Ladhak, Frieda Rong, Hongyu Ren, Huaxiu
  Yao, Jue Wang, Keshav Santhanam, Laurel Orr, Lucia Zheng, Mert Yuksekgonul,
  Mirac Suzgun, Nathan Kim, Neel Guha, Niladri Chatterji, Omar Khattab, Peter
  Henderson, Qian Huang, Ryan Chi, Sang~Michael Xie, Shibani Santurkar, Surya
  Ganguli, Tatsunori Hashimoto, Thomas Icard, Tianyi Zhang, Vishrav Chaudhary,
  William Wang, Xuechen Li, Yifan Mai, Yuhui Zhang, and Yuta Koreeda.
\newblock Holistic evaluation of language models, 2023.
\newblock URL \url{https://arxiv.org/abs/2211.09110}.

\bibitem[Liao and Xiao(2023)]{liao2023rethinking}
Q~Vera Liao and Ziang Xiao.
\newblock Rethinking model evaluation as narrowing the socio-technical gap.
\newblock \emph{arXiv preprint arXiv:2306.03100}, 2023.

\bibitem[Maslej et~al.(2024)Maslej, Fattorini, Perrault, Parli, Reuel,
  Brynjolfsson, Etchemendy, Ligett, Lyons, Manyika, Niebles, Shoham, Wald, and
  Clark]{maslej2024ai}
Nestor Maslej, Loredana Fattorini, Raymond Perrault, Vanessa Parli, Anka Reuel,
  Erik Brynjolfsson, John Etchemendy, Katrina Ligett, Terah Lyons, James
  Manyika, Juan~Carlos Niebles, Yoav Shoham, Russell Wald, and Jack Clark.
\newblock The {AI} index 2024 annual report.
\newblock AI Index Steering Committee, Institute for Human-Centered AI,
  Stanford University, April 2024.

\bibitem[Mouton et~al.(2023)Mouton, Lucas, and Guest]{mouton2023operational}
Christopher~A. Mouton, Caleb Lucas, and Ella Guest.
\newblock \emph{The Operational Risks of AI in Large-Scale Biological Attacks:
  A Red-Team Approach}.
\newblock RAND Corporation, Santa Monica, CA, 2023.
\newblock \doi{10.7249/RRA2977-1}.

\bibitem[{NIST}(2023)]{nist2023artificial}
{NIST}.
\newblock Artificial intelligence risk management framework ({AI RMF 1.0}),
  2023.
\newblock URL \url{https://doi.org/10.6028/NIST.AI.100-1}.
\newblock Accessed: 2024-09-06.

\bibitem[Patwardhan et~al.(2024)Patwardhan, Liu, Markov, Chowdhury, Leet, Cone,
  Maltbie, Huizinga, Wainwright, Jackson, Adler, Casagrande, and
  Madry]{patwardhan2024earlywarning}
Tejal Patwardhan, Kevin Liu, Todor Markov, Neil Chowdhury, Dillon Leet, Natalie
  Cone, Caitlin Maltbie, Joost Huizinga, Carroll Wainwright, Shawn~(Froggi)
  Jackson, Steven Adler, Rocco Casagrande, and Aleksander Madry.
\newblock Building an early warning system for {LLM}-aided biological threat
  creation, 2024.
\newblock URL
  \url{https://openai.com/index/building-an-early-warning-system-for-llm-aided-biological-threat-creation/}.
\newblock Accessed: 2024-09-16.

\bibitem[Phuong et~al.(2024)Phuong, Aitchison, Catt, Cogan, Kaskasoli,
  Krakovna, Lindner, Rahtz, Assael, Hodkinson, Howard, Lieberum, Kumar, Raad,
  Webson, Ho, Lin, Farquhar, Hutter, Deletang, Ruoss, El-Sayed, Brown, Dragan,
  Shah, Dafoe, and Shevlane]{phuong2024evaluating}
Mary Phuong, Matthew Aitchison, Elliot Catt, Sarah Cogan, Alexandre Kaskasoli,
  Victoria Krakovna, David Lindner, Matthew Rahtz, Yannis Assael, Sarah
  Hodkinson, Heidi Howard, Tom Lieberum, Ramana Kumar, Maria~Abi Raad, Albert
  Webson, Lewis Ho, Sharon Lin, Sebastian Farquhar, Marcus Hutter, Gregoire
  Deletang, Anian Ruoss, Seliem El-Sayed, Sasha Brown, Anca Dragan, Rohin Shah,
  Allan Dafoe, and Toby Shevlane.
\newblock Evaluating frontier models for dangerous capabilities, 2024.
\newblock URL \url{https://arxiv.org/abs/2403.13793}.

\bibitem[Roose(2024)]{roose2024ai}
Kevin Roose.
\newblock {A}.{I}. has a measurement problem.
\newblock \emph{The New York Times}, April 2024.
\newblock URL
  \url{https://www.nytimes.com/2024/04/15/technology/ai-models-measurement.html}.
\newblock Accessed: 2024-09-05.

\bibitem[Weidinger et~al.(2023)Weidinger, Rauh, Marchal, Manzini, Hendricks,
  Mateos-Garcia, Bergman, Kay, Griffin, Bariach, Gabriel, Rieser, and
  Isaac]{weidinger2023sociotechnical}
Laura Weidinger, Maribeth Rauh, Nahema Marchal, Arianna Manzini, Lisa~Anne
  Hendricks, Juan Mateos-Garcia, Stevie Bergman, Jackie Kay, Conor Griffin, Ben
  Bariach, Iason Gabriel, Verena Rieser, and William Isaac.
\newblock Sociotechnical safety evaluation of generative {AI} systems.
\newblock \emph{arXiv preprint arXiv:2310.11986}, 2023.

\end{thebibliography}
\bibliographystyle{plainnat}

\end{document}